\documentclass[%
 reprint,
 amsmath,amssymb,
 aps,
]{revtex4-1}

\usepackage{graphicx}% Include figure files
\usepackage{dcolumn}% Align table columns on decimal point
\usepackage{bm}% bold math
\usepackage{multirow}
\usepackage[table,xcdraw]{xcolor}
\usepackage{tikz}
\begin{document}

\preprint{APS/123-QED}

\title{Is Post Selection  Physical: A Device Independent Outlook }% Force line breaks with \\
\author{Anubhav Chaturvedi}
 \affiliation{Center for Computational Natural Sciences and Bioinformatics,\\
 International Institute of Information Technology-Hyderabad, Gachibowli, Hyderabad-500032, India..}%Lines break automatically or can be forced with \\

\author{Tushant Jha, Indranil Chakrabarty}

\affiliation{Center for Security Theory and Algorithmic Research\\
 International Institute of Information Technology-Hyderabad, Gachibowli, Hyderabad, India.
}%

\date{\today}% It is always \today, today,

\begin{abstract}
\noindent The basic motivation behind this work is to raise the question that whether post selection 
can be considered a valid physical transformation (on probability space) or not. We study the consequences of both 
answers set in a device (theory) independent framework, based only on observed statistics. \\ We start with taking up 
post-selection as an assumption (if the answer is \textbf{YES}) and model the same using independent devices governed 
by Boolean functions. We establish analogy between the post selection functions and the general probabilistic games in 
a two party binary input-output scenario. As an observation, we categorize all possible post-selection functions based 
on the effect on a uniform input probability distribution.  We find that post-selection can transform simple no signaling 
probability distributions to signaling. Similarly, solving NP (nondeterministic polynomial time) complete 
problems is easy independent of classical or quantum 
computation (in particular we prove that Post RP (Randomized Polynomial Time) = NP). Finally, we demonstrate an instance of the violation of the pigeon 
hole principle independent of underlying theory. As result of our theory independent modeling we conclude that post-selection 
as an assumption adds  power to the underlying theory. In particular, quantum mechanics benefits 
more with the post-selection assumption, only because it admits a more general set of allowed probabilities as compared to 
the local hidden variable model.  Without the assumption (if the answer is \textbf{NO}) we associate a device independent 
efficiency factor to quantify the cost of post selection. Our study shows that in the real world post-selection is not 
efficient enough to be of any advantage. But from an adversarial perspective it is still of significance. As an application, 
we obtain robust bounds on faking the bell violation (correlation in general) in terms of minimum efficiency required using 
post selection. Here in this work we argue that post-selection as an assumption is  not physical. In the real world  post-selection 
is simply dropping trials based on a pre-decided rule. It makes physical reality appear surprising. However, we suggest 
the use of post-selection with an device independent trial efficiency to avoid anomalous effects.

\end{abstract}

\pacs{Valid PACS appear here}% PACS, the Physics and Astronomy
\maketitle

%\tableofcontents

\section{\label{sec:level1}INTRODUCTION}
\noindent The  mathematical foundation of quantum mechanics was laid down long time back \cite{hilbert1928grundlagen}. 
Einstein questioned the completeness  of quantum mechanics as a fundamental theory through the EPR \cite{einstein1935can} paper in the year 1935.  
He had a strong opinion in support of a deterministic (local and real) explanation to the universe, 
on the other hand intrinsic randomness of QM (lack of reality) was completely antagonistic to his point of view.  However, he never argued against 
the correctness of quantum theory, he only questioned its completeness. He  hinted towards the existence of a 
underlying, "complete" local hidden variable  theory  not very different from classical mechanics  \cite{santos1992critical, ballentine1972einstein}.

\noindent For almost three decades the question on completeness of quantum mechanics was the talking point 
in-spite of increasing experimental evidence \cite{tegmark2001100}, up until the rather revolutionary work by  Bell. 
Bell showed that no local  hidden variable set up can simulate the statistics of quantum entanglement 
\cite{bell1964einstein,home1991bell,guo2002scheme}. 
In effect, the set of quantum probabilities is more general than the set of probabilities admitted by the suggested underlying local hidden variable model. As a consequence much of the research in the last two decades has been focused on entanglement's usefulness as a resource to carry out information processing protocols 
like quantum teleportation~\cite{bennett1993teleporting,adhikari2008teleportation}, cryptography~\cite{gisin2002quantum,lo2012measurement,,ekert2014ultimate,masanes2011secure,acin2007device,hillery1999quantum,colbeck2011private,pironio2010random,chakrabarty2009secret,adhikari2012probabilistic}, 
superdense coding \cite{bennett1992communication}, 
remote state preparation \cite{bennett2001remote,pati2000minimum}, broadcasting of entanglement\cite{chatterjee2014no,adhikari2006broadcasting} 
and many more \cite{sazim2013retrieving}.  Coming back into Bell's scenario, the most important consequence of Bell's work 
and operational outlook on his work was the statistical method of comparing theories, based on observed 
statistics \cite{fry1976experimental,freedman1972experimental}. Bell inequalities \cite{bell1964einstein,fine1982hidden,terhal2000bell} in particular classify correlations and compare theories ( Local hidden variable vs. quantum ) in a 
device independent way, i.e ,without any need to describe the degrees of freedom under study and the measurements that 
are performed.

\noindent  In the recent past we have seen the post selection is an area of interest; and in that 
context we have witnessed various phenomenons where we tend to believe that post selection is primarily responsible for those 
though various other reasons were simultaneously provided \cite{knight1990weak,ferrie2014result,aharonov2014quantum,branciard2011detection}. 
Our capability to solve problems computationally is 
limited by physics \cite{bennett1985fundamental,lloyd2000ultimate,aaronson2004limits}. 
In physics a layer of controversy still surrounds the question whether it is physical to 
comment about nature through post-selected ensembles \cite{aharonov1988result,aharonov1990properties}. Recently the authors of reference \cite{aharonov2014quantum} showed that in a quantum mechanical setting there is violation 
of very basic pigeon hole principle. However, it was not clear that whether quantum mechanics or post selection is responsible for this 
violation.  Using detection loophole Eavesdropper can render QKD protocols 
at lower efficiency unsecured \cite{lydersen2010hacking}. Also any Device-independent (DI) quantum communication will 
require a post-selection loophole-free violation of Bell inequalities \cite{branciard2011detection}.

\noindent The post selection process has an enormous implication in complexity theory. From a complexity theorist 
perspective post-selection is simply conditioning the probability 
space based on the occurrence of an event. Postselection, effectively, allows us to consider only a subset of 
all possible outcomes of an event $E$ by saying that one only considers those outcomes where some other event 
$F$ has taken place.
While P is the class of all problems that can be solved in time polynomial to size of input, NP is the class of problems 
for which there are polynomial-sized proofs for all positive instances that can be verified in time polynomial to size of 
input. And it is one of the major challenges in complexity theory to check whether these classes are equal or not. Between 
these two classes lies RP and BQP. RP is the class of all problems that can be solved with zero probability false positives 
and less than half probability of false negatives. RP defined in this manner, lies between P and NP. BQP (bounded error 
quantum polynomial time) is the class of decision problems solvable by a quantum computer in polynomial time, with an error 
probability of at most 1/3 for all instances. BQP contains P and quantum computers are not known to solve NP-complete 
problems \cite{aaronson2005guest}.
We have seen that working in this new probability space greatly enhances the computation 
capability of a quantum computer by  making 
it as powerful as non-deterministic poly-time Turing machine that accepts if the majority of its paths do (PP). 
In particular the authors of reference \cite{aaronson2005quantum} showed that class of languages decidable by a 
bounded-error polynomial-time quantum computer, if at any time you can measure a qubit that has a nonzero probability of 
being $|1>$, and assume the outcome will be $|1>$, PostBQP is equivalent to PP.

%Post-selected weak 
%measurement is a useful technique to amplify weak physical effects which gives significant improvement in 
%(conditioned) success probabilities of quantum information protocols . Through post-selected ensembles, quantum mechanics 
%seemed to produce rather anomalous effects.    
% However, it was recently shown that the phenomena of post-selected weak measurement and 
%values is observed in a simple classical model. The authors went ahead and called such anomalous effects artifacts of 
%toying with classical statistics and disturbance . 

\noindent In this work we put post-selection through the device independent test. We start with developing the device 
independent framework for two party binary input output scenario.  Next we model post-selection as a device (theory) 
independent transformation on probability space described uniquely by a Boolean function on input and out variables.  
We explore these post-selection functions from two different perspectives. The first one is set in hypothetical world 
where post-selection is a physical (efficient) transformation. We study the relationship between post-selection functions 
and functions governing (non local) games to show that post selection if taken up valid transform can efficiently make 
simple (both quantum and local hidden variable) no signaling probability distribution, signaling. With the help of this 
relationship we categorize all possible post-selection functions. To highlight the importance of the device independent 
modeling, we show that post-selection allows us to solve the NP complete problems efficiently independent of quantum theory.  
In particular we show that post selection strengthens the classical complexity class RP to NP. Further we bring up an 
instance of the violation of pigeon hole principle using only post-selection unlike the claim 
made in the reference \cite{aharonov2014quantum} that quantum mechanics is responsible for such violation. These results 
show that post-selection provides similar nonphysical power to both quantum and local hidden variable models and quantum 
probabilities being a general set admits greater power under post-selection. 
\noindent The second perspective is set in the real world, instead of assuming it as a valid transformation we 
associate a device independent (trial) efficiency factor as the cost of implementation. We study the theory 
independent relationship between the post-selection function, input probability distribution and efficiency 
factor for all the functions used above. We conclude while post-selection provides stupendous power when taken 
up as an assumption, in the real world it is of no advantage due loss of trials.  Finally as an application we 
provide robust bounds over minimum efficiency required to fake quantum correlations using local hidden variable 
correlations as resource, from an adversarial perspective. Which leads to Device Independently secure statistics 
for some observable range of efficiency, reemphasizing the fact that quantum mechanics is more general. While 
these observation clearly suggest that taking post-selection as an assumption is far from physical and on the 
other hand in the real world it distorts physical reality making it anomalous and sometimes surprising (without 
the device independent efficiency associated).

\section{Device Independent Framework: Non Local Games and Post Selection}
\noindent This section lays the prevalent theory independent notions set in binary input-output probability distribution. We define the set of probability distribution under 1.) no-signaling assumption 2.) local hidden variable model 3.) quantum mechanics  in terms of a probability distributions in a two 
party binary input output situation. Some of the well known probability distributions are represented as points on the convex polytope (shown 
in the figure [FIG 1].) In the next subsection we define general probabilistic and non-local games (in particular B-CHSH game). In the next subsection we model theory independent post-selection from two perspectives 1.) Post-selection as an assumption 2.) Post-selection without assumption.
\noindent \subsection{ Device Independent Framework}
\noindent A device independent test is a statistical test wherein we treat the measurement device as a 
black box with classical inputs and outputs. Let Alice and Bob be two spatially separated parties. Alice (Bob) has 
a device with binary input  $x (y) \in \{0,1\}$ and binary output $u (v) \in \{0,1\}$. For each trial Alice and 
Bob randomly chose the input $x (y)$ such that $P(x=0)=P(y=0)=\frac{1}{2}$ (Experimental Free Will \cite{conway2006free}). They receive 
the output $u (v)$. They collect the statistics of several trials to construct individual  $P(u|x)$ and  $P(v|y)$, 
 the joint  $P(u,v|x,y)$  probability distributions using communication where,
\begin{equation}
P(u|x)=P(u|x,y)=\sum_vP(u,v|x,y).
\end{equation}
\noindent  The first equality is because of a fundamental bound on spatially separated communication called no- signaling which 
tells us that output on one side is independent of what is given as input on the other side. 
For a complete no-signaling correlation $P(u,v|x,y)$ we will have
\begin{equation}
I(A:B)=I(x:y,v)=0, 
\end{equation}
\begin{equation}
I(B:A)=I(y:x,u)=0. 
\end{equation}
\noindent  
where mutual information between Alice's independent input $x$ and Bob's system ($y,v$) $I(x:y,v)=H(x)-H(x|y,v)$ and $H(x),H(x|y,v)$ are Shannon's entropy and Shannon's conditional entropy. No-signaling is a fundamental principle and forms a convex polytope in the conditional probability distribution space with eight vertices's, within which the following probability distributions lie. For a two dimensional realization of the polytope (see FIG 1.).\\

\begin{figure}[htp]
\centering
\includegraphics[scale=0.45]{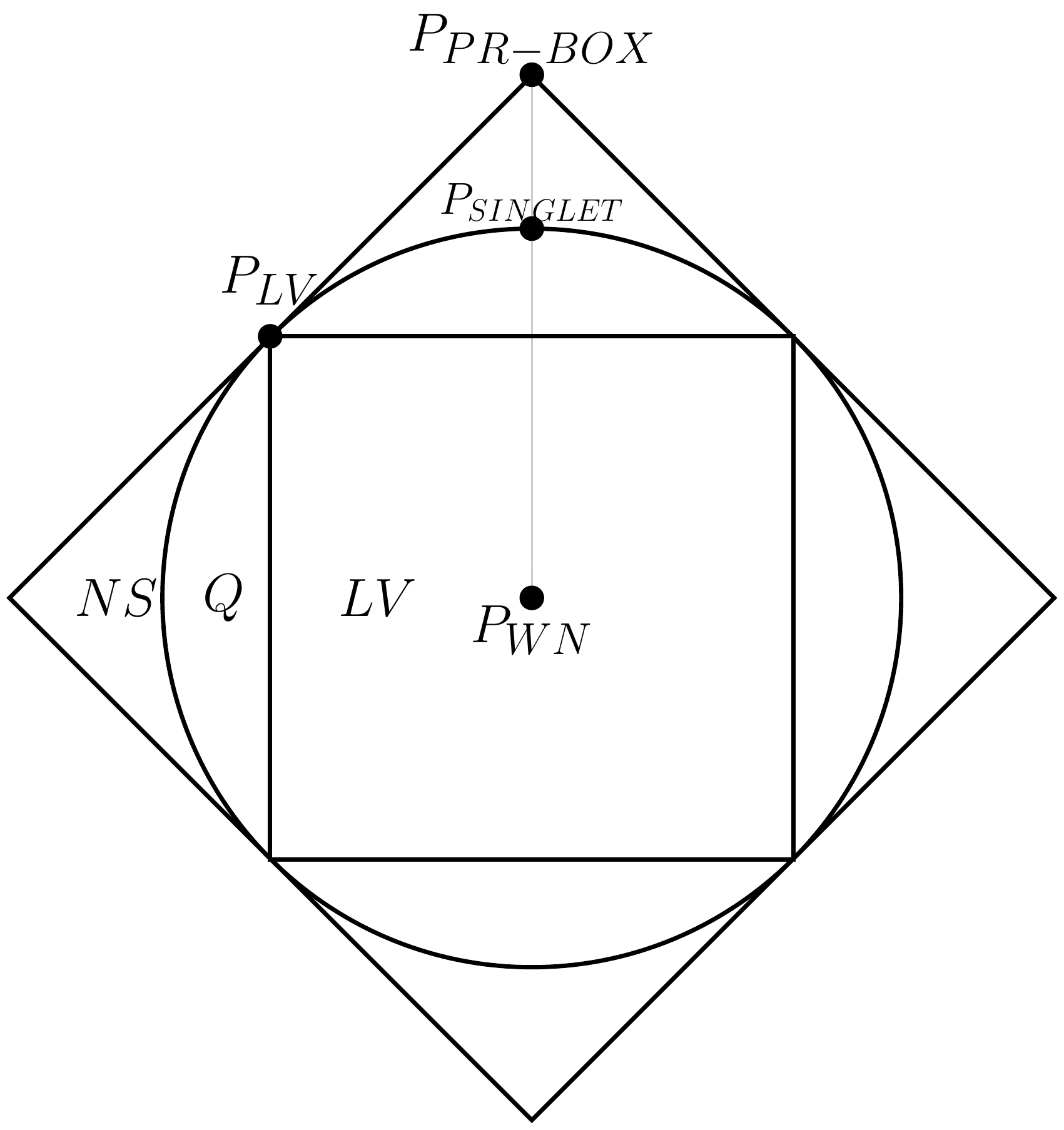}
\caption{No signaling Polytope: The outer square demarcates the no signaling ($NS$) probability distribution with PR boxes 
as the extremal points ($P_{PR-BOX}$). The inner circle is the boundary of the convex set admitted by quantum mechanics ($Q$) where $P_{SINGLET}$ 
is the probability distribution of the singlet. The inner square 
is the boundary of probability distributions admitted Local Hidden Variable Model $LV$ with white noise $P_{WN}$ at the center.}
\end{figure}

% \usepackage{multirow}

\noindent \textbf{White Noise:}\\

\noindent The center point of this convex polytope (see FIG 1) is the white noise. The conditional probability distribution of the outputs 
$u$, $v$ given the inputs $x$ and $y$  i.e.  $P_{WN}(u,v|x,y)$ in the  TABLE I:\\
\begin{table}[h]
\label{tbl:WN}
\begin{tabular}{|c|c|c|c|c|c|}
\hline
\multicolumn{2}{|c|}{}                                    & \multicolumn{2}{c|}{$x=0$}                                                    & \multicolumn{2}{c|}{$x=1$}                                                    \\ \cline{3-6} 
\multicolumn{2}{|c|}{\multirow{-2}{*}{$P_{WN}(u,v|x,y)$}} & $u=0$                                 & $u=1$                                 & $u=0$                                 & $u=1$                                 \\ \hline
                                     & $v=0$              & $\frac{1}{4}$ & $\frac{1}{4}$                         &$\frac{1}{4}$ & $\frac{1}{4}$                         \\ \cline{2-6} 
\multirow{-2}{*}{$y=0$}              & $v=1$              & $\frac{1}{4}$                         & $\frac{1}{4}$ & $\frac{1}{4}$                         & $\frac{1}{4}$ \\ \hline
                                     & $v=0$              & $\frac{1}{4}$ & $\frac{1}{4}$                         & $\frac{1}{4}$                         &$\frac{1}{4}$ \\ \cline{2-6} 
\multirow{-2}{*}{$y=1$}              & $v=1$              & $\frac{1}{4}$                         & $\frac{1}{4}$ & $\frac{1}{4}$ & $\frac{1}{4}$                         \\ \hline
\end{tabular}
\caption{ White Noise: Probability Distribution of $P_{WN}(u,v|x,y)$ with $P_{WN}(f_{B-CHSH}=0)=\frac{1}{2}$ for binary inputs $x,y$ and outputs $u,v$}
\end{table}
\noindent  
$P_{WN}$ is the uniform probability distribution which is the center for the Local Hidden Variable convex polytope and Quantum convex set.\\

\noindent \textbf{Local Hidden Variable Model:}\\

\noindent The idea of local hidden variable model for any 
hidden variable $\lambda$, (pre-established agreement) is based on  assumptions: 
(1) Measurement Independence: $P(\lambda|x,y)=P(\lambda)$, (2) 
Outcome Independence: $P(u,v|x,y,\lambda)=P(u|x,\lambda)P(v|y,\lambda)$. Combining these two conditions we get,
$P_{LV}(u,v|x,y)=\sum_{\lambda}P(\lambda)P(u|x,\lambda)P(v|y,\lambda)$.
\noindent The point $P_{LV}$  on the no signaling polytope is given in the figure 1 (FIG 1). The probability distribution of a local hidden variable 
model is shown in TABLE II.\\
\begin{table}[h]
\begin{tabular}{|c|c|c|c|c|c|}
\hline
\multicolumn{2}{|c|}{}                                    & \multicolumn{2}{c|}{$x=0$}                                & \multicolumn{2}{c|}{$x=1$}                                \\ \cline{3-6} 
\multicolumn{2}{|c|}{\multirow{-2}{*}{$P_{LV}(u,v|x,y)$}} & $u=0$                       & $u=1$                       & $u=0$                       & $u=1$                       \\ \hline
                                     & $v=0$              & $1$ & $0$                         & $1$                         & $0$ \\ \cline{2-6} 
\multirow{-2}{*}{$y=0$}              & $v=1$              & $0$                         &$0$ & $0$ & $0$                         \\ \hline
                                     & $v=0$              & $1$ & $0$                         & $1$                         & $0$ \\ \cline{2-6} 
\multirow{-2}{*}{$y=1$}              & $v=1$              & $0$                         &$0$ & $0$ & $0$                         \\ \hline
\end{tabular}
\caption{ Local Hidden Variable Model: Probability Distribution of $P_{LV}(u,v|x,y)$ with $P_{LV}(f_{B-CHSH}=0)=\frac{3}{4}$ for binary inputs $x,y$ and outputs $u,v$}
\end{table}

\noindent \textbf{Quantum Mechanics:}\\

\noindent %As we know that quantum mechanics can never be replaced by any local hidden variable model. 
Any $P(u,v|x,y)$ is said to belong to the set pf quantum mechanical probability distributions $P_{Q}(u,v|x,y)$ if one can find a 
quantum state $\rho\in H$ (where $H$ is the Hilbert space) and measurements $M^x=\{E^x_u|u\in\{0,1\}\}$,$M^y=\{E^y_v|v\in\{0,1\}\}$ such that,
$P_Q(u,v|x,y)=trace(\rho E^x_uE^y_v)$ holds.  $P_Q$ forms a convex set with infinite external points. For the singlet quantum state and bell measurements we have the 
probability distribution as shown in TABLE III.\\

\begin{table}[h]
\begin{tabular}{|c|c|c|c|c|c|}
\hline
\multicolumn{2}{|c|}{}                                         & \multicolumn{2}{c|}{$x=0$}                                                                          & \multicolumn{2}{c|}{$x=1$}                                                                          \\ \cline{3-6} 
\multicolumn{2}{|c|}{\multirow{-2}{*}{$P_{SINGLET}(u,v|x,y)$}} & $u=0$                                            & $u=1$                                            & $u=0$                                            & $u=1$                                            \\ \hline
                                        & $v=0$                &$\frac{2+\sqrt[]{2}}{8}$ & $\frac{2-\sqrt[]{2}}{8}$                         & $\frac{2+\sqrt[]{2}}{8}$ & $\frac{2-\sqrt[]{2}}{8}$                         \\ \cline{2-6} 
\multirow{-2}{*}{$y=0$}                 & $v=1$                & $\frac{2-\sqrt[]{2}}{8}$                         & $\frac{2+\sqrt[]{2}}{8}$ & $\frac{2-\sqrt[]{2}}{8}$                         & $\frac{2+\sqrt[]{2}}{8}$ \\ \hline
                                        & $v=0$                & $\frac{2+\sqrt[]{2}}{8}$ & $\frac{2-\sqrt[]{2}}{8}$                         & $\frac{2-\sqrt[]{2}}{8}$                         & $\frac{2+\sqrt[]{2}}{8}$ \\ \cline{2-6} 
\multirow{-2}{*}{$y=1$}                 & $v=1$                & $\frac{2-\sqrt[]{2}}{8}$                         & $\frac{2+\sqrt[]{2}}{8}$ & $\frac{2+\sqrt[]{2}}{8}$ & $\frac{2-\sqrt[]{2}}{8}$                         \\ \hline
\end{tabular}
\caption{Singlet: Probability Distribution of $P_{SINGLET}(u,v|x,y)$ with $P_{singlet}(f_{B-CHSH}=0)=\frac{2+\sqrt[]{2}}{4}$ for binary inputs $x,y$ and outputs $u,v$}
\end{table}

\noindent \textbf{Popescu Rohlich Box:}\\

\noindent The eight vertices of the no signaling polytope are functionally similar to the $P_{PR-BOX}(u,v|x,y)$ and together form the external 
 points of the polytope. Recently there has been a lot of research aimed at finding physical principles that do not allow $P_{PR-BOX}$ to exist in nature \cite{pawlowski2009information,yang2011quantum}. The probability distribution of $P_{PR-BOX}(u,v|x,y)$ is shown in TABLE IV.
\begin{table}[h]
\begin{tabular}{|c|c|c|c|c|c|}
\hline
\multicolumn{2}{|c|}{}                                        & \multicolumn{2}{c|}{$x=0$}                                                    & \multicolumn{2}{c|}{$x=1$}                                                    \\ \cline{3-6} 
\multicolumn{2}{|c|}{\multirow{-2}{*}{$P_{PR-BOX}(u,v|x,y)$}} & $u=0$                                 & $u=1$                                 & $u=0$                                 & $u=1$                                 \\ \hline
                                       & $v=0$                & $\frac{1}{2}$ & $0$                                   & $\frac{1}{2}$ & $0$                                   \\ \cline{2-6} 
\multirow{-2}{*}{$y=0$}                & $v=1$                & $0$                                   & $\frac{1}{2}$ & $0$                                   & $\frac{1}{2}$ \\ \hline
                                       & $v=0$                & $\frac{1}{2}$ & $0$                                   & $0$                                   & $\frac{1}{2}$ \\ \cline{2-6} 
\multirow{-2}{*}{$y=1$}                & $v=1$                & $0$                                   & $\frac{1}{2}$ &$\frac{1}{2}$ & $0$                                   \\ \hline
\end{tabular}
\caption{ PR Box: Probability Distribution of $P_{PR-BOX}(u,v|x,y)$ with $P_{PR-BOX}(f_{B-CHSH}=0)=1$ for binary inputs $x,y$ and outputs $u,v$}
\end{table}

\noindent \subsection{Non-local games}
\noindent By  a non-local game we 
refer to one of the task 
in the family of cooperative tasks (general probabilistic games) for a team of several remote
players, where every player is randomly assigned an input by a verifier. Each of these players then chooses
one out of a set of possible outputs and sends it to the verifier. The verifier then determines the success 
probability according to a predefined condition $f=0$ where the function is given by, $f:\{0,1\}^4\rightarrow\{0,1\}$ .
The players know the winning condition and may coordinate
a joint strategy. In bipartite situation like in our case, the joint strategy is given by the probability 
distribution $P(u,v|x,y)$.
The success probability of the task (say $f$) given a strategy is $P(u,v|x,y)$, 
$P(f=0)=\frac{\sum_{f=0}P(u,v|x.y)}{4}$.
\noindent A team making use of quantum correlations (shared entanglement) is
said to employ a \textquotedblleft{}quantum strategy\textquotedblright{},
whereas if not, is said to employ a \textquotedblleft{}classical strategy\textquotedblright{}.\\
 
\noindent \textit{Definition 1:} A non local game is one whose success probability distinguishes between probability distributions admitted by local hidden variable theory from 
the ones admitted by only quantum theory(or in general no-signaling ). The winning probability of a non-local game must follow,
$max_{LV}(P(f=0))<max_Q(P(f=0))\leq1$.\\

\noindent One such game is the B-CHSH game given by the function $f_{B-CHSH}(u,v,x,y)=u.v\oplus x \oplus y$. The bell 
inequality can be written in terms of the winning probability associated with  the game,
$max_{LV}(P(f_{B-CHSH}=0))=\frac{3}{4}$. This gives us a facet of LV polytope. It is maximally violated by an entangled quantum state, 
$max_{Q}(P(f_{B-CHSH}=0))=P_{SINGLET}(f_{B-CHSH}=0)=\frac{2+\sqrt[]{2}}{4}$. The PR-BOXs are super quantum no-signaling strategies which maximally violate Bell inequality,
$max_{NS}(P_{B-CHSH}(f=0))=P_{PR-BOX}(f_{B-CHSH}=0)=1$. Any probability distribution lying on the line joining $P_{PR-BOX}$
and $P_{WN}$ is given by the form, $P_{B-CHSH}(c)=cP_{PR-BOX}+(1-c)P_{WN}$.
Here $c\in\{0,1\}$ is a convex coefficient or simply classical mixing parameter. In TABLE V we have given the probability 
distribution of $P_{B-CHSH}(c)(u,v|x,y)$ for input $x,y$ and output $u,v$

\begin{table}[h]
\begin{tabular}{|c|c|c|c|c|c|}
\hline
\multicolumn{2}{|c|}{}                                         & \multicolumn{2}{c|}{$x=0$}                                                        & \multicolumn{2}{c|}{$x=1$}                                                        \\ \cline{3-6} 
\multicolumn{2}{|c|}{\multirow{-2}{*}{$P_{B-CHSH}(c)(u,v|x,y)$}} & $u=0$                                   & $u=1$                                   & $u=0$                                   & $u=1$                                   \\ \hline
                                        & $v=0$                & $\frac{1+c}{2}$ & $\frac{1-c}{2}$                         & $\frac{1+c}{2}$ & $\frac{1-c}{2}$                         \\ \cline{2-6} 
\multirow{-2}{*}{$y=0$}                 & $v=1$                & $\frac{1-c}{2}$                         & $\frac{1+c}{2}$ & $\frac{1-c}{2}$                         & $\frac{1+c}{2}$ \\ \hline
                                        & $v=0$                & $\frac{1+c}{2}$ & $\frac{1-c}{2}$                         & $\frac{1-c}{2}$                         & $\frac{1+c}{2}$ \\ \cline{2-6} 
\multirow{-2}{*}{$y=1$}                 & $v=1$                & $\frac{1-c}{2}$                         & $\frac{1+c}{2}$ & $\frac{1+c}{2}$ & $\frac{1-c}{2}$                         \\ \hline
\end{tabular}
\caption{ $P_{B-CHSH}(c)$: Probability Distribution of $cP_{PR-BOX}(u,v|x,y)+(1-c)P_{WN}(u,v|x,y)$ for binary inputs $x,y$ and outputs $u,v$}.
\end{table}
\noindent This strategy when used for B-CHSH has the success probability,
$P_{B-CHSH}(f_{B-CHSH}=0)=c(1)+(1-c)(\frac{1}{2})=\frac{(1+c)}{2}$. 
%For each value of $P(f=0)$ there is a 
%corresponding $P_{B-CHSH}$ which lies on the B-CHSH line i.e. a convex 
%combination $P_{WN}$ and $P_{PR-BOX}$. For example we get the singlet correlation at, $c_{SINGLET}=\frac{1}{\sqrt[]{2}}$.

\noindent \subsection{Post-selection }

\noindent \subsubsection{Post-selection as an assumption}

\noindent \textbf{Assumption:} Post-selection is an  efficient transformation on probability space.\\

\noindent To post select for an event $E$, the probability of some other event F changes from $P[F]$ to the conditional 
probability $P[F|E]$. The assumption implies we can (somehow) instantaneously perform post-selection without loss 
of efficiency. Any event $E$ in a classical (input-output) setup can represented by a condition $f(u,v,x,y)=0$ 
where $f$ is post-selection governing Boolean function. \\
 
\noindent \textit{Definition 2.} A Post-Selection Device (PSD(f)) is a device which takes in input probability distribution 
$P_{in}(u,v|x,y)$ and accepts the trial if $f(u,v,x,y)=0$. Output 
probability distribution then simply becomes, $P_{out}(u,v|x,y)$=$P_{in}(u,v|x,y,f=0)$.\\

We can pre-select the input probability distribution $P_{in}(u,v|x,y)$ which simply specifies $P[F]$. 
Pre-selection(-paration) is the theory dependent part of our skeleton. As in one can only prepare a $P_{in}(u,v|x,y)$ 
which is allowed by the theory.\\ 
%We can now define the characteristic probability distribution $P_{f-BOX}$,\\

\noindent \textit{Definition 3.} $P_{f-BOX}$ associated with a $PSD(f)$ is
\begin{equation}
P_{f-BOX}(u,v|x,y)=P_{WN}(u,v|x,y,f=0)
\end{equation}

\noindent In fact, all application of post-selection can be modeled with the help of two steps : pre- and post-selection. \\ 

\noindent \textbf{Properties:}\\

\noindent Next we introduce two important properties of post selection which we are going to use later. \\

\noindent \textit{a) Sequential application and orthogonal functions.}\\

\noindent For a discrete probability space, $P[F|E] = \frac{P(F \wedge E)}{P[E]}$, and thus for post-selection to be 
well defined we require that $P[E]>0$. We start with $P_{WN}$ simply for the fact that for all 
functions $f:\{0,1\}^4\rightarrow\{0,1\}$,  $P_{WN}(f=0)>0$ except $f=1$. Two functions $f$ and $f^1$  are called 
orthogonal if they cannot be applied sequentially to $P_{WN}$. For example, if $P_{f-BOX}(f^1=0)=0$ then
one cannot apply post-selection function $f^1 $ after $f$ and vice-versa. 
The probability distributions $P_{f-BOX}$ and $P_{f^1-BOX}$ are orthogonal that is one 
cannot be post-selected from other using any $f$ (see FiG 2).\\  

\begin{figure}[htp]
\centering
\includegraphics[scale=0.6]{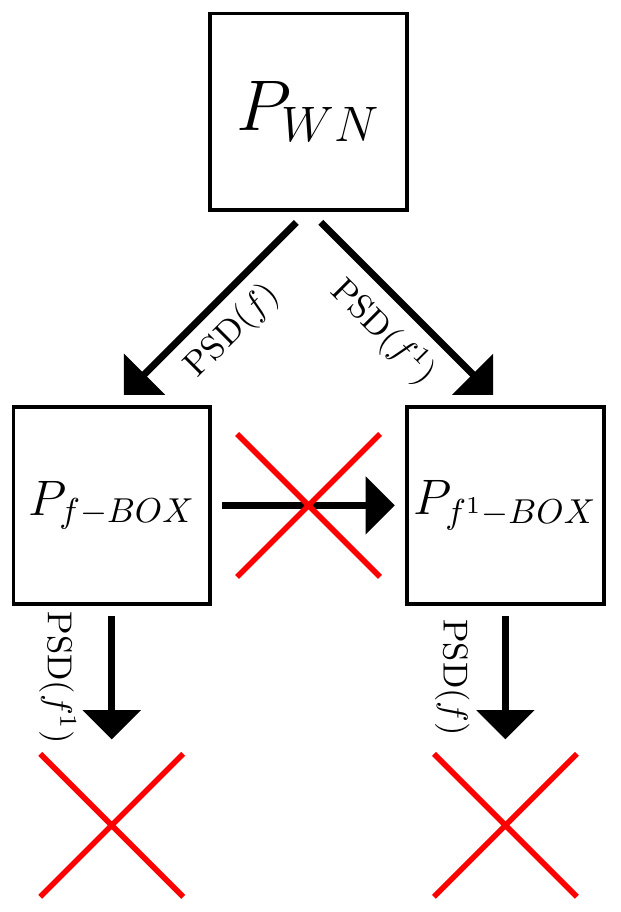}
\caption{Orthogonal functions: If $f$ and $f^1$ are othogonal 1.) one cannot post-select them sequential with $P_{WN}$ as input and 2.) $P_{f-BOX}$ can't be transformed into $P_{f^1-BOX}$ using any PSD.} 
\end{figure}

\noindent \textit{b) Boolean compliments: }\\ 

\noindent $P_{WN}$ can be prepared in many ways, which are indistinguishable from a device independent perspective. $P_{WN}$ is the center of the no-signaling polytope. So it could be broken down into infinite pairs of \textbf{'complementary'} correlations $P_{A,B},P'_{A,B}$ such that,
\begin{equation} \label{12}
P_{WN}=\frac{P_{A,B}+P'_{A,B}}{2}.
\end{equation}
 
\noindent If two functions $f$ and $f^1$are Boolean compliments that is $f=f^1\oplus 1$  then,
1) they must be orthogonal and,  2) they must produce complimentary correlations as output to 
$P_{WN}=\frac{P_{f-BOX}+P_{f^1-BOX}}{2}$. We can independently and simultaneously apply $f$ and $f^1$  as in principle we could have two post selection devices 
applied to $P_{WN}$ such that one accepts when $f=0$ and the other when $f=1$ (see FIG 3.).

\begin{figure}[htp]
\centering
\includegraphics[scale=0.6]{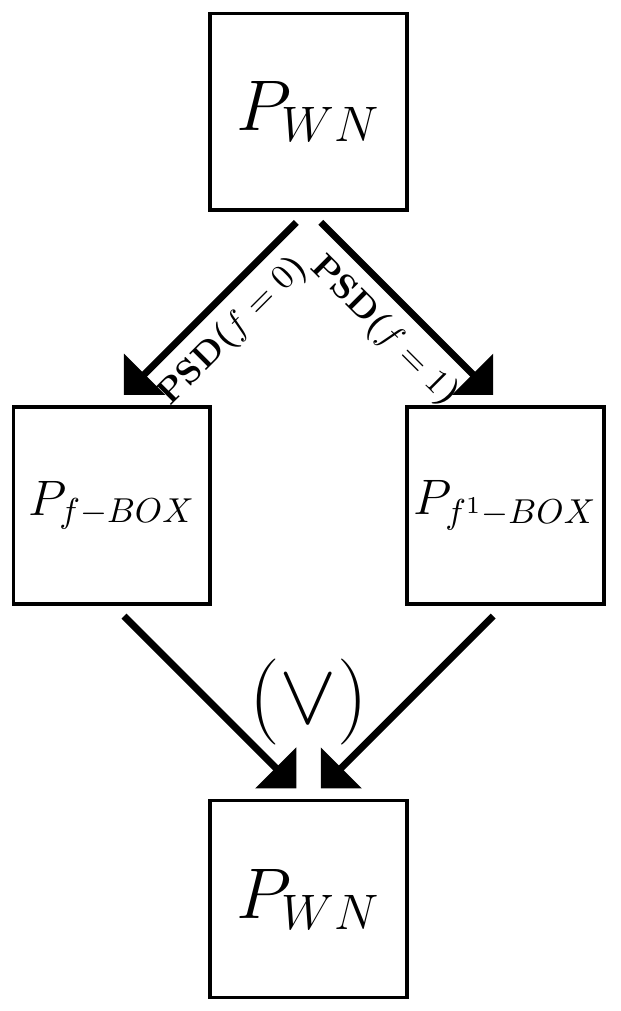}
\caption{Simultaneous application of PSD($f$) and PSD($f^1$), $f^1=f\oplus 1$. The outputs could again be mixed to form $P_{WN}$ that is $f\oplus f^1\oplus 1=0$ always holds.} 
\end{figure}

\noindent \subsubsection{Post-selection without the assumption.}

\noindent In this we do not consider in general post selection to be efficient transformation 
in probability space. Post-selection in today's world is basically a trial by trial evaluation of 
the input probability distribution wherein one simply accepts when $f=0$ (say) and ignores when $f=1$. 
It is easy to see that Post selection requires substantial amount of communication to get the input outputs 
of the two spatially separated  parties to evaluate a  Boolean function $f(x,y,u,v)$. In this context let us 
define an efficiency factor associated with the success probability of the post selection function given the communication required.\\

\noindent \textit{Definition 4.} The efficiency of applying Post Selection function  on the input probability distribution 
$P_{in}$ resulting in $P_{out}$ is given by $\eta^{P^{in}}_f\in (0,1)$. This efficiency factor is 
given by the success probability of the function $f$ to take the value $0$ i.e $\eta^{P^{in}}_f=P(f=0)$. 
Note that this efficiency is independent of the theory governing the boxes and only depends on $P_{in}$ and $f$.

\noindent \section{Implication of Post-selection as an assumption.}
\noindent In this section we show that if we assume post selection to be a efficient transformation on the 
probability space, we observe the following surprising implications. This includes, 1) transforming no signaling probability 
distribution to signaling probability distribution, 2) solving NP Complete Problems 3) violation of Pigeon Hole Principle. 
 
\subsection{ No signaling Theories to Signaling Theories via post-selection.}
\noindent In this subsection we show that on application of fully efficient post selection function we can change no 
signaling probability distribution to signaling probability distribution.We white noise as 
input probability distribution because $P_{WN}(f=0)>0$ for all $f(x,y,u,v)$ and change it to signaling probability distribution. Similarly many other no signaling 
probability distributions that lie on the convex polytope are converted to signaling probability distributions
on application of simple post selection described by Boolean functions. This inter convertibility is best represented by the schematic 
diagram given by the figure.\\

\noindent Let us consider the case where we take the input probability distribution as the white noise
(probability distribution given in table 1) $P_{in}=P_{WN}$. 
One can obtain on application of a PSD($f$) a $P_{out}$, such that $P_{out}(f=0)=\sum_{f=0}P_{out}(u,v|x,y)=1$. The point $P_{out}$ is also referred to as $P_{f-BOX}$. 
In general for every Boolean function  there exist a classical 
input output task (game) and a PSD($f$) that
takes $P_{WN}$ to the correlation $P_{f-BOX}$ tailor made to win the task (with success probability = 1).  
This simply implies that using post-selection one can win all such tasks completely and violate all the physical principles associate with such tasks. 
The success probability of the associated the game $f$ can also be alternatively reported as the "projection" 
of a point on the line joining  $P_{f-BOX}$ and $P_{WN}$. On the basis of this observation we can formally 
categorize the set of $f$ (post-selection functions).\\

\noindent \textbf{Signaling/No-signaling:}\\

\noindent \textit{Definition 5.} A function $f$ is called one-way(Alice-Bob) signaling iff,
\begin{equation}
I_{f-BOX}(A:B)>0.
\end{equation}
%or example $f_{sig1}:v\oplus x$ shown in Table VI is a trivial Alice to Bob signaling function with $I(A:B)=1$.\\

\noindent \textit{Definition 6.} A function $f$ is called one-way(Bob-Alice) signaling iff,
\begin{equation}
I_{f-BOX}(B:A)>0.
\end{equation}
%$f_{sig2}:  u\oplus y$ shown in the table VII is an example of Bob to Alice signaling function with mutual information 
%$I(A:B)=1$.\\

\noindent \textit{Definition 7.}
If both of the condition are met i.e  $I_{f-BOX}(A:B)>0$ and $I_{f-BOX}(B:A)>0$  then $f$ is called both 
side signaling. 
%In table VIII we provide an example of a function $f$ for which both side signaling is happening 
%with both the mutual information $I(A:B), I(B:A)$ being equal to $1$.\\

\noindent \textit{Definition 7.} We say that a function is no-signaling when the following conditions are simultaneously met.
\begin{equation}
I_{f-BOX}(A:B)=0,
\end{equation}
\begin{equation}
I_{f-BOX}(B:A)=0.
\end{equation}

\noindent \textbf{Local/non-local :}\\

\noindent \textit{Definition 8.}
We say that a function $f$ is local if,
\begin{equation}
P_{f-BOX}(x.y=u\oplus y)\leq \frac{3}{4}.
\end{equation}
%All no-signaling functions independent Alice's(Bob's) input $x$($y$) trivially local. 
\\
  
\noindent \textit{Definition 9.} We say that a function $f$ is non-local if,
\begin{equation}
P_{f-BOX}(x.y=u\oplus y)>\frac{3}{4}.
\end{equation}

\noindent  Interestingly there are only 5 non-local no-signaling functions. One of them is B-CHSH. Other  functions are similar upto renaming to $f_{NL}$. 
The probability distribution for this non local box $P_{f_{NL}-BOX}(u,v|x,y)$ is shown in the TABLE VI.  
\begin{table}[h]
\begin{tabular}{|c|c|c|c|c|c|}
\hline
\multicolumn{2}{|c|}{}                                            & \multicolumn{2}{c|}{$x=0$}                                                                        & \multicolumn{2}{c|}{$x=1$}                                                                        \\ \cline{3-6} 
\multicolumn{2}{|c|}{\multirow{-2}{*}{$P_{f_{NL}-BOX}(u,v|x,y)$}} & $u=0$                                           & $u=1$                                           & $u=0$                                           & $u=1$                                           \\ \hline
                                         & $v=0$                  & $\frac{1}{4}$                                   & $\frac{1}{4}$                                   & $\frac{1}{2}$                                   &  \\ \cline{2-6} 
\multirow{-2}{*}{$y=0$}                  & $v=1$                  & $\frac{1}{4}$                                   & $\frac{1}{4}$                                   & & $\frac{1}{2}$                                   \\ \hline
                                         & $v=0$                  & $\frac{1}{2}$                                   &  &  & $\frac{1}{2}$                                   \\ \cline{2-6} 
\multirow{-2}{*}{$y=1$}                  & $v=1$                  &  & $\frac{1}{2}$                                   & $\frac{1}{2}$                                   &  \\ \hline
\end{tabular}
\caption{Non Local Box: Probability distribution of $P_{f_{NL}-BOX}(u,v|x,y)$ with $P_{f_{NL}-BOX}(f_{B-CHSH}=0)=\frac{7}{8} $for binary inputs $x$ and $y$ and outputs $u$ and $v$. The empty boxes signifies the positions where $f_{NL}=1$ and can be taken as $0$. 
}
\end{table}
%It is important to note that the tasks based on these functions are non-local games. 
%Interestingly there are only 5 non-local no-signaling 
%functions. One of them is B-CHSH.\\

\noindent In the FIG 4, we show a part of no signaling polytope and the transformation of the initial probability 
distribution $P_{in}=P_{WN}$ to various probability distributions with the application of post selection functions 
$f_{sig1}$, $f_{sig2}$, $f_{sig}$, $f_{B-CHSH}$, $f_{CTC}$. We take specific examples: 
$f_{sig1}:v\oplus x$, $f_{sig2}:  u\oplus y$, $f_{sig}: (x\oplus v\oplus1).(u\oplus y\oplus1) \oplus 1$, $f_{B-CHSH}:x.y \oplus u\oplus v$, $f_{CTC}:y\oplus v$. \\

\begin{figure}[htp]
\centering
\includegraphics[scale=0.7]{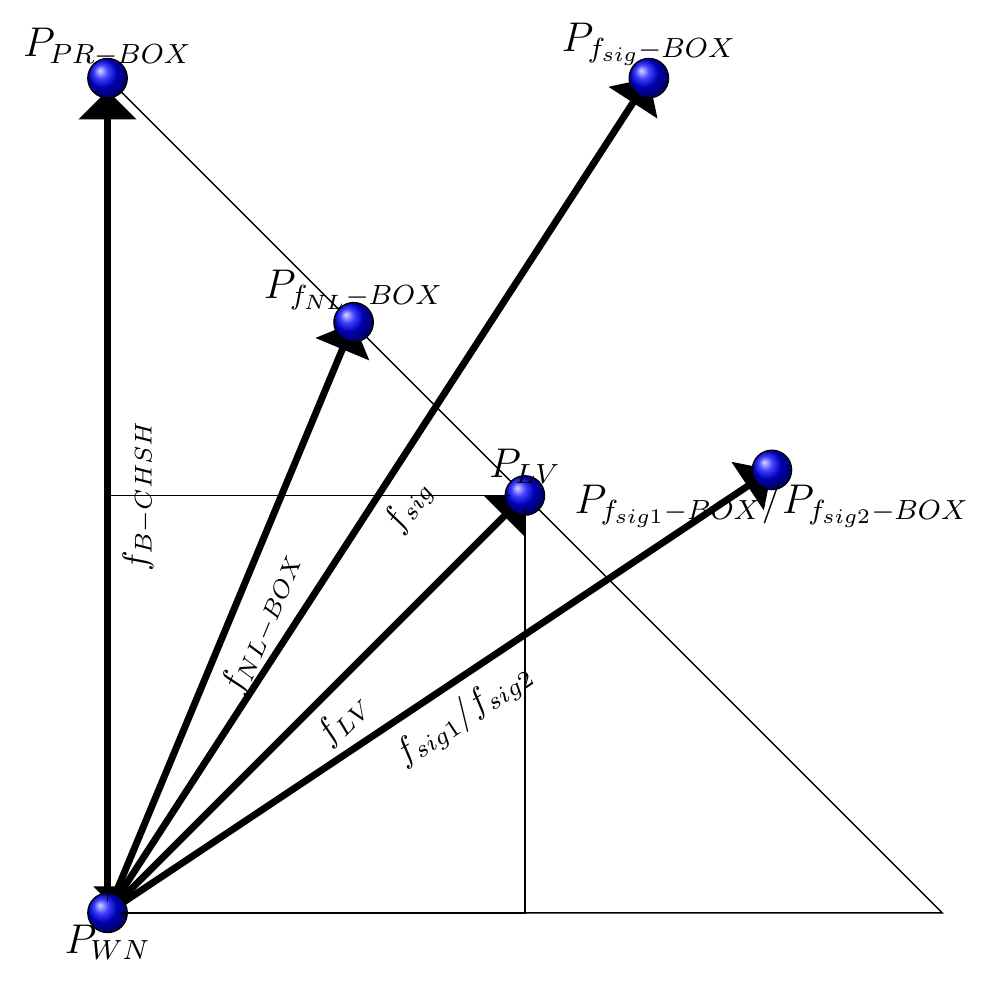}
\caption{Transformation of Probability Distribution: In this schematic diagram the transformation of the 
initial probability distribution $P_{WN}$ to different output probability distribution on application of different post 
selection functions $f: f_{sig1}, f_{sig2}, f_{sig}, f_{B-CHSH}, f_{CTC}$ is shown.}.
\label{2}
\end{figure}

%We say $f$ is a signaling function if either one of the two or both of 
%the following conditions are true,
%\begin{equation}
%I_{f-BOX}(A:B)>0
%\end{equation}

\noindent In the TABLE VII we enlist down the signaling and no signaling possibilities (by evaluating 
$I(A:B)$ and $I(B:A)$ ). These post selection functions are $f_{sig1}$, $f_{sig2}$, $f_{sig}$, $f_{B-CHSH}$, $f_{CTC}$. 
The input probability distribution are $P_{WN}$, $P_{LV}$, $P_{SINGLET}$, $P_{PR-BOX}$, $P_{f_{NL}-BOX}$ (all no signaling probability distribution). We calculate each of 
the mutual information $I(A:B)$ and $I(B:A)$. In a nut shell 
this table gives a holistic view how the post selection when applied on an input probability distribution changes 
no signaling probability distributions to signaling probability distributions. \\

\begin{table}[h]
\label{tbl:VII}
\begin{tabular}{|c|c|c|c|}
\hline
\textbf{$f$}                  & \textbf{$P_in$}  & \textbf{$I(A:B)$} & \textbf{$I(B:A)$} \\ \hline
\multirow{5}{*}{$f_{sig1}$}   & $P_{WN}$         & $1$               & $0$               \\ \cline{2-4}
                              & $P_{LV}$         & $1$               & $0$               \\ \cline{2-4}
                              & $P_{SINGLET}$    & $1$               & $0.394553$        \\ \cline{2-4}
                              & $P_{PR-BOX}$     & $1$               & $\frac{1}{2}$     \\ \cline{2-4}
                              & $P_{f_{NL}-BOX}$ & $1$               & $\frac{1}{2}$     \\ \hline
\multirow{5}{*}{$f_{sig2}$}   & $P_{WN}$         & $0$               & $1$               \\ \cline{2-4}
                              & $P_{LV}$         & $0$               & $1$               \\ \cline{2-4}
                              & $P_{SINGLET}$    & $0.394553$        & $1$               \\ \cline{2-4}
                              & $P_{PR-BOX}$     & $\frac{1}{2}$     & $1$               \\ \cline{2-4}
                              & $P_{f_{NL}-BOX}$ & $\frac{1}{2}$     & $1$               \\ \hline
\multirow{5}{*}{$f_{sig}$}    & $P_{WN}$         & $1$               & $1$               \\ \cline{2-4}
                              & $P_{LV}$         & $1$               & $1$               \\ \cline{2-4}
                              & $P_{SINGLET}$    & $1$               & $1$               \\ \cline{2-4}
                              & $P_{PR-BOX}$     & $1$               & $1$               \\ \cline{2-4}
                              & $P_{f_{NL}-BOX}$ & $1$               & $1$               \\ \hline
\multirow{5}{*}{$f_{B-CHSH}$} & $P_{WN}$         & $0$               & $0$               \\ \cline{2-4}
                              & $P_{LV}$         & $0$               & $0$               \\ \cline{2-4}
                              & $P_{SINGLET}$    & $0$               & $0$               \\ \cline{2-4}
                              & $P_{PR-BOX}$     & $0$               & $0$               \\ \cline{2-4}
                              & $P_{f_{NL}-BOX}$ & $0$               & $0$               \\ \hline
\multirow{5}{*}{$f_{CTC}$}    & $P_{WN}$         & $0$               & $0$               \\ \cline{2-4}
                              & $P_{LV}$         & $0$               & $0$               \\ \cline{2-4}
                              & $P_{SINGLET}$    & $0.394553$        & $0$               \\ \cline{2-4}
                              & $P_{PR-BOX}$     & $\frac{1}{2}$      & $0$               \\ \cline{2-4}
                              & $P_{f_{NL}-BOX}$ & $\frac{1}{4}$      & $0$               \\ \hline
\end{tabular}
\caption{Application of the Post Selection functions $f$: $f_{sig1}$, $f_{sig2}$, $f_{sig}$, $f_{B-CHSH}$, $f_{CTC}$ 
on Input No signaling probability distributions $P_{in}$: $P_{WN}$, $P_{LV}$, $P_{SINGLET}$, $P_{PR-BOX}$, $P_{f_{NL}-BOX}$   }
\end{table}

\noindent \textbf{Few Specific Examples :}\\

\noindent Apart from providing the previous table where we have shown the transition of a no signaling probability distributions 
to a signaling probability distributions on application of post selection functions; here also we provide few specific examples 
in TABLES VII, VIII, IX,X, XI with much more detailing.\\

\begin{itemize}

\item In TABLE VIII we provide an example where Alice to Bob 
signaling is taking place i.e $I(A:B)=1$. In this case we take the input probability distribution as the 
white noise $P_{WN}$ and the post selection function as $f_{sig1}$.

\item In TABLE IX we provide an example where Bob to Alice signaling is taking 
place  i.e $I(B:A)=1$. Here also we take the input probability distribution as the white noise 
$P_{WN}$ and this time the post selection function is $f_{sig2}$.

\item In TABLE X we show the case where both way signaling is possible with the same input probability distribution 
$P_{WN}$ and the post signaling function as $f_{sig}$. 

\item In the next TABLE XI we take the input probability distribution as a convex combination of PR box and white noise 
i.e $P_{in}= cP_{PR-BOX}+(1-c)P_{WN}$ where $0<c<1$. In this case the post selection function is  $f_{CTC}$ which when 
applied to $P_{in}$ we find the mutual information as 
$I(A:B)=\frac{1}{2}(1-\frac{1+c}{2}\log{\frac{1+c}{2}}+\frac{1-c}{2}\log{\frac{1-c}{2}})$. But $I(B:A)=0$ implying that 
there is only one-side signaling. Its interesting 
to note that it takes all other correlations on the line joining $P_{WN}$ and $P_{PR-BOX}$ to signaling. 

\item In TABLE XII we take the input probability distribution as $P_{in}= cP_{NL-BOX}+(1-c)P_{WN}$ where $0<c<1$. The post selection 
function $f_{CTC}$ is same as the previous case. The mutual information in this case is given by
$I(A:B)=\frac{1}{4}(1-\frac{1+c}{2}\log{\frac{1+c}{2}}+\frac{1-c}{2}\log{\frac{1-c}{2}})$.   
\end{itemize}

\begin{table}[t]
\begin{tabular}{|c|c|c|c|c|c|}
\hline
\multicolumn{2}{|c|}{}                                       & \multicolumn{2}{c|}{$x=0$}                                                                                                      & \multicolumn{2}{c|}{$x=1$}                                                                                                      \\ \cline{3-6} 
\multicolumn{2}{|c|}{\multirow{-2}{*}{$P_{f-BOX}(u,v|x,y)$}} & $u=0$                                                          & $u=1$                                                          & $u=0$                                                          & $u=1$                                                          \\ \hline
                                       & $v=0$               & $\frac{1}{2}$                                                  & $\frac{1}{2}$                                                  &                & \               \\ \cline{2-6} 
\multirow{-2}{*}{$y=0$}                & $v=1$               &                                       &                                       & $\frac{1}{2}$                                                  & $\frac{1}{2}$                                                  \\ \hline
                                       & $v=0$               & $\frac{1}{2}$                                                  & $\frac{1}{2}$                                                  &   &  \\ \cline{2-6} 
\multirow{-2}{*}{$y=1$}                & $v=1$               &   &   & $\frac{1}{2}$                                                  & $\frac{1}{2}$                                                \\ \hline
\end{tabular}
\caption{Application of post selection function $f_{sig1}$ on $P_{WN}$: Alice to Bob signaling, $I(A:B)=1$.}
\end{table}
%\noindent This condition represents Alice to Bob signaling. For example $f:v\oplus x$  trivial Alice to Bob signaling function. 
%\begin{equation}
%I_{f-BOX}(B:A)>0
%\end{equation}

\begin{table}[h]
\begin{tabular}{|c|c|c|c|c|c|}
\hline
\multicolumn{2}{|c|}{}                                       & \multicolumn{2}{c|}{$x=0$}                                                                        & \multicolumn{2}{c|}{$x=1$}                                                                        \\ \cline{3-6} 
\multicolumn{2}{|c|}{\multirow{-2}{*}{$P_{f-BOX}(u,v|x,y)$}} & $u=0$                                           & $u=1$                                           & $u=0$                                           & $u=1$                                           \\ \hline
                                       & $v=0$               & $\frac{1}{2}$                                   &  & $\frac{1}{2}$                                   &  \\ \cline{2-6} 
\multirow{-2}{*}{$y=0$}                & $v=1$               & $\frac{1}{2}$                                   &  & $\frac{1}{2}$                                   &  \\ \hline
                                       & $v=0$               &  & $\frac{1}{2}$                                   & & $\frac{1}{2}$                                   \\ \cline{2-6} 
\multirow{-2}{*}{$y=1$}                & $v=1$               &  & $\frac{1}{2}$                                   & & $\frac{1}{2}$                                   \\ \hline
\end{tabular}
\caption{Application of post selection function $f_{sig2}$ on $P_{WN}$: Bob to Alice signaling, $I(B:A)=1$.}
\end{table}
%\noindent This condition represents Bob to Alice signaling. For example $f:  u\oplus y$ is a trivial Bob to Alice signaling function.
%If both of the condition are met for example $f: (x\oplus v\oplus1).(u\oplus y\oplus1) \oplus 1$.
\begin{table}[h]
\begin{tabular}{|c|c|c|c|c|c|}
\hline
\multicolumn{2}{|c|}{}                                       & \multicolumn{2}{c|}{$x=0$}                                                                        & \multicolumn{2}{c|}{$x=1$}                                                                        \\ \cline{3-6} 
\multicolumn{2}{|c|}{\multirow{-2}{*}{$P_{f-BOX}(u,v|x,y)$}} & $u=0$                                           & $u=1$                                           & $u=0$                                           & $u=1$                                           \\ \hline
                                       & $v=0$               & $1$                                   &  &            &  \\ \cline{2-6} 
\multirow{-2}{*}{$y=0$}                & $v=1$               &                         &  & $1$                                   & \\ \hline
                                       & $v=0$               &  & $1$                                   & &                         \\ \cline{2-6} 
\multirow{-2}{*}{$y=1$}                & $v=1$               & &                        &  & $1$                                   \\ \hline
\end{tabular}
\caption{Application of post selection function $f_{sig}$ on $P_{WN}$: Both side signaling, $I(A:B)=1$ and $I(B:A)=1$.}
\end{table}

%\noindent Closed Time Like curves can take PR-BOX to signaling when one party has access to a world line 
%along closed time-like curve  \cite{chakrabarty2014ctc}. Let Alice and Bob share a correlation lying on line 
%joining $P_{PR-BOX}$ and $P_WN$. Say Bob get access to world line along CTC. Let $f^{CTC}=y\oplus v$ (feedback loop) 
%be the post-selection that represents the action of CTC. 
\begin{table}[h]
\begin{tabular}{|c|c|c|c|c|c|}
\hline
\multicolumn{2}{|c|}{}                                             & \multicolumn{2}{c|}{$x=0$}                                                                                                      & \multicolumn{2}{c|}{$x=1$}                                                                                            \\ \cline{3-6} 
\multicolumn{2}{|c|}{\multirow{-2}{*}{$P_{f^{CTC}-BOX}(u,v|x,y)$}} & $u=0$                                                          & $u=1$                                                          &  $u=0$ & $u=1$           \\ \hline
                                          & $v=0$                  & $\frac{1+c}{2}$                        &  $\frac{1-c}{2}$ & $\frac{1+c}{2}$              &  $\frac{1-c}{2}$ \\ \cline{2-6} 
\multirow{-2}{*}{$y=0$}                   & $v=1$                  &                 &                                        &      &                                        \\ \hline
                                          & $v=0$                  &                                       &                &       &                                      \\ \cline{2-6} 
\multirow{-2}{*}{$y=1$}                   & $v=1$                  & $\frac{1-c}{2}$ & $\frac{1+c}{2}$                        & $\frac{1+c}{2}$              &  $\frac{1-c}{2}$ \\ \hline
\end{tabular}
\caption{Application of post selection function $f_{CTC}$ on $P_{in}= cP_{PR-BOX}+(1-c)P_{WN}$: Alice to Bob signaling.}
\end{table}

%\noindent We find that the mutual information 
%\begin{equation}
%I(A:B)=\frac{1}{2}(1-\frac{1+p}{2}\log{\frac{1+p}{2}}+\frac{1-p}{2}\log{\frac{1-p}{2}})
%\end{equation} and the amount increases with the increasing value of $P(f^{B-CHSH}=0)$. Similarly the effect on the other line joining $P_{BOX}$ and $P_{WN}$ is, 
\begin{table}[h]
\begin{tabular}{|c|c|c|c|c|c|}
\hline
\multicolumn{2}{|c|}{}                                             & \multicolumn{2}{c|}{$x=0$}                                                                                                    & \multicolumn{2}{c|}{$x=1$}                                                                                            \\ \cline{3-6} 
\multicolumn{2}{|c|}{\multirow{-2}{*}{$P_{f^{CTC}-BOX}(u,v|x,y)$}} & $u=0$                                                          & $u=1$                                                        &  $u=0$ &  $u=1$          \\ \hline
                                          & $v=0$                  & $\frac{1}{2}$                          &  $\frac{1}{2}$ & $\frac{1+p}{2}$              &  $\frac{1-p}{2}$ \\ \cline{2-6} 
\multirow{-2}{*}{$y=0$}                   & $v=1$                  &                 &                                      &       &                                       \\ \hline
                                          & $v=0$                  &                                       &             &       &                                        \\ \cline{2-6} 
\multirow{-2}{*}{$y=1$}                   & $v=1$                  &  $\frac{1-p}{2}$ & $\frac{1+p}{2}$                      & $\frac{1+p}{2}$              &  $\frac{1-p}{2}$ \\ \hline
\end{tabular}
\caption{Application of post selection function $f_{CTC}$ on $P_{in}= cP_{NL-BOX}+(1-c)P_{WN}$: Alice to Bob signaling.}
\end{table}

\subsection{Post-selection, RP and NP-Completeness}
\noindent $NP$ stands for \textit{“non deterministic polynomial time,”} a term going back to the roots of
complexity theory \cite{papadimitriou2003computational}. Intuitively, it means that a solution to any search problem can be found
and verified in polynomial time by a special (and quite unrealistic) sort of algorithm, called a
non deterministic algorithm. Such an algorithm has the power of guessing correctly at every
step. Incidentally, the original definition of $NP$ (and its most common usage to this day) was
not as a class of search problems  but as a class of decision problems. In other words $NP$ is set of all decision 
problems which can be verified, but not necessarily be solved, in polynomial time.\\ 
\begin{figure}[htp]
\centering
\includegraphics[scale=0.45]{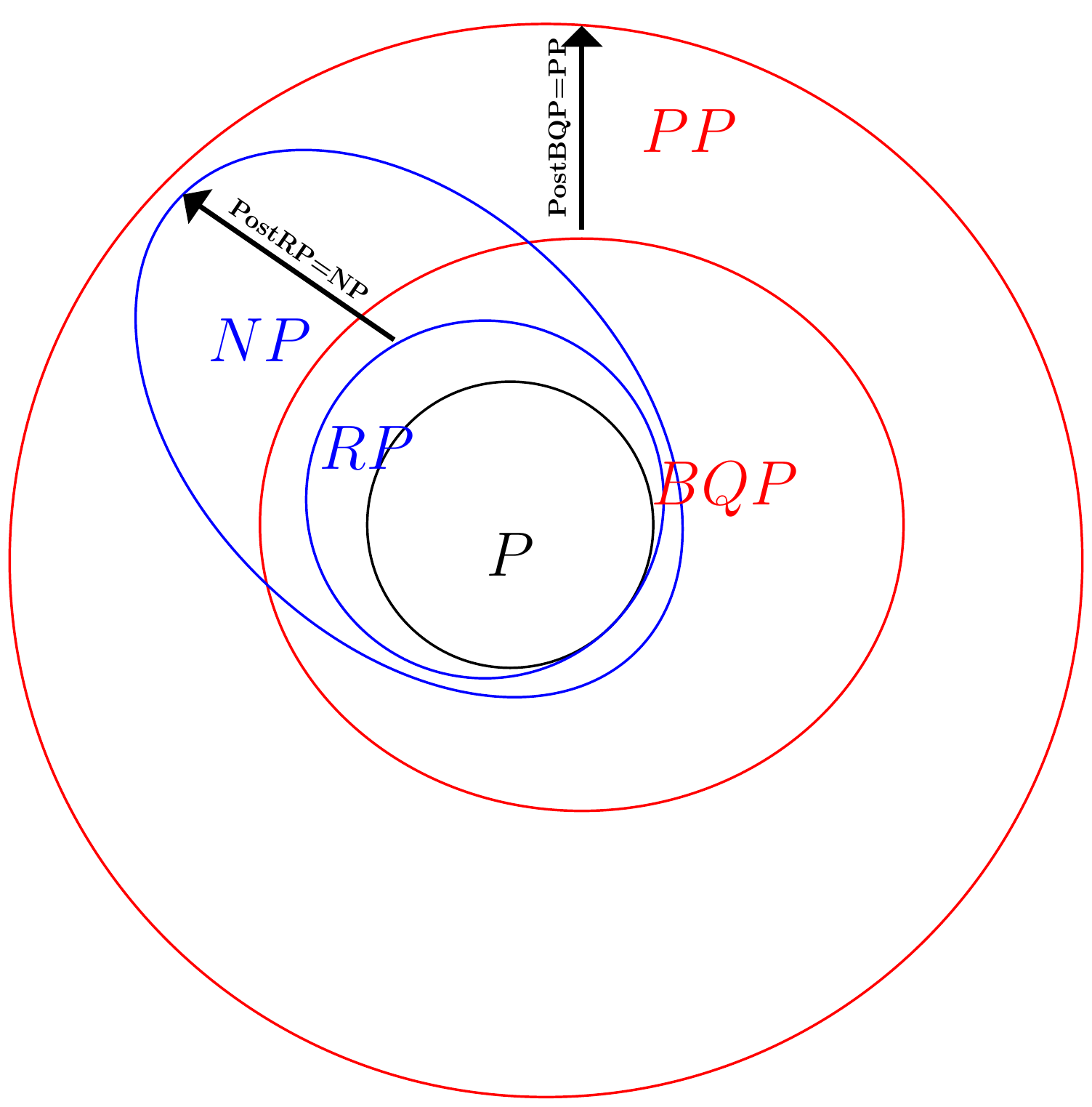}
\caption{This figure describes the relationship between the complexity classes $P,RP,BQP,NP,PP$. $PostBQP=PP$(red) and $PostRP=NP$(blue).}
\end{figure}

\noindent \textit{Definition 10.}
A language $L$ is said to belong to class $NP$ iff for every $x \in L$, there exists a $y$, 
such that $\mid y \mid \le p(\mid x \mid)$, for some polynomial 
$p$, and $L^\prime = \lbrace (x,y) : x \in L \rbrace$ can be decided in polynomial time.\\

\noindent In other words, a problem is considered to be in the class $NP$, if for every true instance of the problem, there exists a proof of the answer, with polynomial bounded length, such that given the input and the proof, the proof can be verified in polynomial time.
In complexity theory, the canonical $NP-complete$ problem, to which all problems of $NP$ can be reduced to, is $3SAT$, i.e. deciding whether a Boolean CNF formula with every $m$ clauses of size 3, over $n$ variables, is satisfy able or not. Therefore, if we can solve $3SAT$ in any framework, we can solve any NP problem in that framework.
A common classical probability framework is defined by $RP$, which is the class of all problems which can be solved by a probabilistic Turi	ng machine in polynomial time, such that error for No-instances is zero, and error for Yes-instances is less than $\frac{1}{2}$.
While it is not known whether $RP$ is equal to $NP$ or not, it is known that $RP \subseteq NP$. However, in this section, we consider the post selection version of $RP$, and discuss its equivalence to $NP$.\\

\noindent \textit{Definition 11.}
\noindent A language $L$ is considered to be in class $PostRP$ iff there exists a probabilistic Turing Machine $M$, that for any input $x$, returns output $Q$ and a flag (on which one post selects) $P$ such that
\begin{enumerate}
\item $Pr(P=1)>0$,
\item For $x \notin L$, $Pr(Q=1\mid P=1) = 0$,
\item For $x \in L$, $Pr(Q=1\mid P=1) \ge \frac{1}{2}$.
\end{enumerate}

\noindent We now consider the randomness in the probabilistic Turing Machine explicitly, 
to make some observations about the nature of the language $L$.
The machine $M$ can be interpreted to compute, in polynomial time, two functions $P$ and $Q$, given 
original input $x$ and a string of polynomially many random bits $r$ as inputs.
Therefore, by converting the nonzero probabilities from the definition of Post RP to existential statements 
on the random string $r$, we get the following corresponding assertions:\\

\begin{enumerate}
\item $\forall x: \exists r: P(x,r)=1$,
\item $\forall x: x \notin L \implies \nexists r: Q(x,r)=1 \wedge P(x,r)=1$,
\item $\forall x: x \in L \implies \exists r: Q(x,r)=1 \wedge P(x,r)=1$.
\end{enumerate}

\noindent Therefore, a proof scheme for such a problem directly follows from its probabilistic Turing machine. By assuming the random string $r$ as a proof of membership, a verifier can simply compute $P$ and $Q$ in polynomial time, and check whether both are equal to $1$.
This scheme results in a membership proof, since
\begin{itemize}
\item For non-members, no such $r$ exists, hence no proof exists.
\item For members (i.e. Yes-instances), there does exist a proof that can be verified in polynomial time.
\end{itemize}

\noindent Now we show that $NP \subseteq PostRP$ by constructing a probabilistic Turing Machine $M$ that can solve $3SAT$ in $PostRP$.
We define $M$ as
\begin{enumerate}
\item Guess variable assignment $\sigma$, uniformly at random,
\item Check whether $\sigma$ satisfies the formula $\Phi$, and assign Q=1 if it does
a) if No (i.e. Q=0), then assign P=1 with probability $\alpha$ b) 
if Yes (i.e. Q=1), then assign P=1 with probability $2^n \times \alpha$.
\end{enumerate}

\noindent For a 3 SAT formula $\Phi$, let $0 \le s \le 2^n$ denote the number of satisfying solutions. Thus $\Phi \in 3SAT \iff s > 0$. Since the machine $M$ guesses an assignment and checks if it is a satisfying assignment, we know that if $s=0$ then $Pr(Q=1) = 0$.
Therefore we note that if $s=0$ and $\Phi \notin 3SAT$, then $Pr(P=1)$ is governed only by the case where $Q=0$, and thus equals $\alpha$. Therefore, $Pr(P=1)>0$. Also, since $Pr(Q=1)=0$, therefore $Pr(Q=1 \bar P=1)=0$.

\noindent While, if $s > 0$ and $\Phi \in 3SAT$, then $Pr(P=1) = \frac{(2^n - s)\alpha + s 2^n \alpha}{2^n} > 0$. Also since the assignments are guessed uniformly at random, $Pr(Q=1) = \frac{s}{2^n} > 0$. Therefore,
\begin{eqnarray}
Pr(Q=1 \mid P=1)&&
= \frac{Pr(P=1 \mid Q=1)Pr(Q=1)}{Pr(P=1)} {}\nonumber\\&&
= \frac{2^n \alpha \times s}{(2^n - s)\alpha + s 2^n \alpha}{}\nonumber\\&&
= \frac{2^n s}{2^n + (2^n - 1)s} {}\nonumber\\&&
> 1/2.%since s>=1
\end{eqnarray}

\noindent Thus we show that the machine $M$ solves $3SAT$ as per \textit{Definitions 10, 11}, thereby making $3SAT \in PostRP$, and it follows from $NP-completeness$ of $3SAT$ that $NP \subseteq PostRP$.
 We, thus, complete our proof of $NP = PostRP$. Post selection strengthens RP up till NP.

\subsection{Violating Pigeon-Hole principle}

\noindent One of the most simple yet fascinating principle of nature is the pigeonhole principle which captures the very 
essence of counting. In a way this principle tells us that if we put three pi-
pigeons in two pigeonholes at least two of the pi-
geons end up in the same hole. In other way round this implies that always there is a non-zero probability of 
finding \textbf{any} two pigeons in the same box. Recently in a work it was shown that in quantum mechanics this 
is not true. They found instances when three quantum particles are put in two boxes, yet no two particles are
in the same box. Here in this section we show that post selection violates the pigeon hole principle independent 
of the theoretical setting.\\

\noindent \textbf{Pigeon Hole principle:} If you put three pigeons in two pigeonholes at least two of the pigeons end up in 
the same hole.\\

\noindent \textbf{Violation of Pigeon Hole principle:} Finding an instance when three pigeons are put in two box 
where no two pigeons are in the same box.\\

\noindent \textbf{Claim:} Our claim is to show that post selection is alone responsible for the violation of the principle 
independent of any theoretical setting.\\

\subsubsection{Modeling and the skeleton.}
\noindent We treat the pigeons as general probability distributions or black boxes with fixed inputs and outputs. 
Let us take three pigeons $A,B,C$. Here we are concerned about only two properties of the pigeons: 1)\textbf{color}(Red or Blue) and 
2)\textbf{hole} (Left or Right).  Here two questions are allowed to ask to each pigeon. These questions are 
denoted by $x,y,z\in\{0,1\}$ for three pigeons $A,B,C$ respectively. Consequently they are allowed to give  
binary answers(outputs) $u,v,w\in\{0,1\}$ respectively. 
If input $x=0$ we  need the answer to say the color of $A$ which could be  $u=0$ (say Red) or  $u=1$ (say Blue) 
and similarly  if $x=1$ we want to know in which hole $A$ is i.e. $u=0$ (say Left) and $u=1$ (say Right). 
Similar questions and answers also hold for other two pigeons $B$ and $C$. 

\noindent These boxes are completely described by the associated probabilities  $P_{A,B,C}(u,v,w|x,y,z)$. To obtain 
individual probabilities like $P_A(u|x)$  and pair wise probabilities like $P_{A<B}(u,v|x,y)$ one can simply 
trace(sum) out other systems. For example,
\begin{equation}
P_{A,B}(u,v|x,y)=\frac{\sum_{w,z}{P_{A,B,C}(u,v,w|x,y,z)}}{2}.
\end{equation}

\subsubsection{ The Pre-selection.}
\noindent Here we preselect the initial probability distribution as the uniform probability distribution of white noise 
i.e, $P_{A,B,C}(u,v,w|x,y,z)=\frac{1}{8}$  for all $u,v,w,x,y,z\in\{0,1\}$. 
Now we make our basic assumption,\\

\noindent \textbf {Assumption: The pigeons are same upto renaming.}\\

\noindent This allows us to reduce the number to two, $P_{A,B}=P_{WN}$ such that $P(u,v|x,y)=\frac{1}{4}$  for 
all $u,v,x,y\in\{0,1\}$.

\subsubsection{ The Post-selection.}
\noindent Let $f_{final}=(x\oplus 1).(y\oplus 1).(z\oplus 1).(u\oplus 1).(v\oplus 1).(w\oplus 1)\oplus 1$  be the PSD governing function. 
The output probability distribution $P_{out}$ is simply $P(u=0,v=0,w=0|x=0,y=0,z=0)=1$and  $P(u,v,w|x,y,z)=0$ for all other cases.

\noindent Notice that this function selects the pigeons with the same color, it has nothing to do 
with the hole in which they are present and we start with a White-Noise distribution, therefore the pigeon 
hole principle is still valid.
%\begin{table}[h]
%\begin{tabular}{|c|c|c|c|c|c|}
%\hline
%\multicolumn{2}{|c|}{}                                       & \multicolumn{2}{c|}{$x=0$}                                                                        & \multicolumn{2}{c|}{$x=1$}                                                                                  \\ \cline{3-6} 
%\multicolumn{2}{|c|}{\multirow{-2}{*}{$P_{f_{final}-BOX}(u,v|x,y)$}} & $u=0$                                           & $u=1$                                           &  $u=0$ &  $u=1$ \\ \hline
%                                       & $v=0$               & 1                       &  &                             &       \\ \cline{2-6} 
%\multirow{-2}{*}{$y=0$}                & $v=1$               &  &                         &       &                             \\ \hline
%                                       & $v=0$               &                         &  &       &                              \\ \cline{2-6} 
%\multirow{-2}{*}{$y=1$}                & $v=1$               &  &                         &                            &       \\ \hline
%\end{tabular}
%\caption{$P_{f_{final}-BOX}$}
%\end{table}

\subsubsection{The question.}
\noindent We pre-select and post-select the same no-violation probability distributions as described above. We need only find a 
possible path where the probability of the pigeons being in the same hole is zero. We question the path that 
could have been taken in between. Our question is whether a function $f_1$ could have been applied in between 
or not? In other words, looking only at the final post-selection we need to find whether $P_{WN}$ could have 
passed through $P_{f_1-BOX}$ or not in world with the assumption. 
\begin{table}[h]
\begin{tabular}{|c|c|c|c|c|c|}
\hline
\multicolumn{2}{|c|}{}                                         & \multicolumn{2}{c|}{$x=0$}                                                                                                  & \multicolumn{2}{c|}{$x=1$}                                                                                                  \\ \cline{3-6} 
\multicolumn{2}{|c|}{\multirow{-2}{*}{$P_{f_1-BOX}(u,v|x,y)$}} & $u=0$                                                        & $u=1$                                                        &  $u=0$         &  $u=1$         \\ \hline
                                        & $v=0$                &                                     &  $\frac{1}{2}$ &                                      & $\frac{1}{2}$ \\ \cline{2-6} 
\multirow{-2}{*}{$y=0$}                 & $v=1$                &  $\frac{1}{2}$ &                                      &  $\frac{1}{2}$ &                                      \\ \hline
                                        & $v=0$                &  $\frac{1}{2}$                        &              &  $\frac{1}{2}$ &                                     \\ \cline{2-6} 
\multirow{-2}{*}{$y=1$}                 & $v=1$                &              & $\frac{1}{2}$                        &                                     &  $\frac{1}{2}$ \\ \hline
\end{tabular}
\caption{$P_{f_{1}-BOX}$ and $P_{f_{1}-BOX}(f_{B-CHSH}=\frac{3}{4})$}
\end{table}
\noindent Notice $f_{final}$ and $f_1$ are orthogonal PS functions. Let $f_2$ be a Boolean complement of $f_1$ and  
can be applied simultaneously.

\begin{table}[h]
\begin{tabular}{|c|c|c|c|c|c|}
\hline
\multicolumn{2}{|c|}{}                                         & \multicolumn{2}{c|}{$x=0$}                                                                                                  & \multicolumn{2}{c|}{$x=1$}                                                                                  \\ \cline{3-6} 
\multicolumn{2}{|c|}{\multirow{-2}{*}{$P_{f_2-BOX}(u,v|x,y)$}} & $u=0$                                                        & $u=1$                                                        &  $u=0$ &  $u=1$ \\ \hline
                                        & $v=0$                & $\frac{1}{2}$                        &              & $\frac{1}{2}$                &      \\ \cline{2-6} 
\multirow{-2}{*}{$y=0$}                 & $v=1$                &              & $\frac{1}{2}$                        &       & $\frac{1}{2}$                \\ \hline
                                        & $v=0$                &                                      &  $\frac{1}{2}$ &      & $\frac{1}{2}$                \\ \cline{2-6} 
\multirow{-2}{*}{$y=1$}                 & $v=1$                &  $\frac{1}{2}$ &                                      & $\frac{1}{2}$                &       \\ \hline
\end{tabular}

\caption{$P_{f_{2}-BOX}$ and $P_{f_{2}-BOX}(f_{B-CHSH}=\frac{3}{4})$}
\end{table}

\noindent Notice $f_2$ (see FiG 6) is not orthogonal to $f_{final}$ and therefore is a valid path. Notice after 
application of $f_2$ pigeons would necessarily be in different holes. So using post-selection one 
can violate the pigeon hole between to non-violating states.
\begin{figure}[htp]
\centering
\includegraphics[scale=0.6]{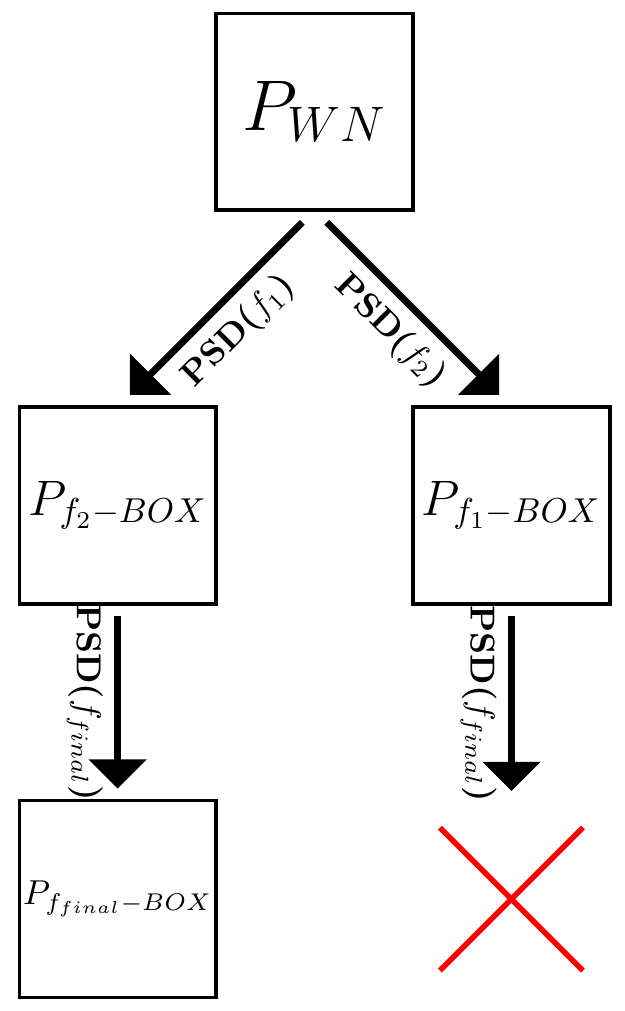}
\caption{ This figure describes theory independent violation of the pigeon hole principle. $f_1,f_{final}$ are orthogonal as $P_{{f_1}-BOX}(f_{final}=0)=0$. So one cannot post-select $f_{final}$ after $f_1$. On the other hand $f_2=f_1\oplus 1$, implying $f_2,f_{final}$ are not orthogonal and hence a valid path in which any two pigeons must be in different holes. }
\end{figure}

\section{ Post-selection without the assumption.}

\noindent In this section, we associate an efficiency factor $\eta^f_{P_{in}}$ with each of these transformations (described by Boolean function $f$) for a given input probability 
distribution $P_{in}$ . We consider the examples  used in the previous section and calculate the efficiency factor. In the next subsection we discuss the role of post-selection from an adversarial perspective and find out the robust bounds on 
maximum efficiency required for simulating non local correlations from an adversarial perspective.    

\subsection{ Evaluating Efficiency Factor}
\noindent In a world without the assumption the loss of trial(efficiency) is the key factor. 
In TABLE XV we provide the device independent efficiency ($\eta^{P_{in}}_f$) for a given post selection 
function $f$ and input probability distribution $P_{in}$.\\

\begin{table}[h]
\begin{tabular}{|c|c|c|}
\hline
\textbf{$f$}                  & \textbf{$P_in$}  & \textbf{$\eta^{P_{in}}_f$} \\ \hline
\multirow{5}{*}{$f_{sig1}$}   & $P_{WN}$         & $\frac{1}{2}$              \\ \cline{2-3}
                              & $P_{LV}$         & $\frac{1}{2}$              \\ \cline{2-3}
                              & $P_{SINGLET}$    & $\frac{1}{2}$              \\ \cline{2-3}
                              & $P_{PR-BOX}$     & $\frac{1}{2}$              \\ \cline{2-3}
                              & $P_{f_{NL}-BOX}$ & $\frac{1}{2}$              \\ \hline
\multirow{5}{*}{$f_{sig2}$}   & $P_{WN}$         & $\frac{1}{2}$              \\ \cline{2-3}
                              & $P_{LV}$         & $\frac{1}{2}$              \\ \cline{2-3}
                              & $P_{SINGLET}$    & $\frac{1}{2}$              \\ \cline{2-3}
                              & $P_{PR-BOX}$     & $\frac{1}{2}$              \\ \cline{2-3}
                              & $P_{f_{NL}-BOX}$ & $\frac{1}{2}$              \\ \hline
\multirow{5}{*}{$f_{sig}$}    & $P_{WN}$         & $\frac{1}{4}$              \\ \cline{2-3}
                              & $P_{LV}$         & $\frac{1}{4}$              \\ \cline{2-3}
                              & $P_{SINGLET}$    & $0.1616$              \\ \cline{2-3}
                              & $P_{PR-BOX}$     & $\frac{1}{8}$              \\ \cline{2-3}
                              & $P_{f_{NL}-BOX}$ & $\frac{1}{16}$             \\ \hline
\multirow{5}{*}{$f_{B-CHSH}$} & $P_{WN}$         & $\frac{1}{2}$              \\ \cline{2-3}
                              & $P_{LV}$         & $\frac{3}{4}$           \\ \cline{2-3}
                              & $P_{SINGLET}$    & $\frac{2+\sqrt{2}}{4}$      \\ \cline{2-3}
                              & $P_{PR-BOX}$     & $1$                        \\ \cline{2-3}
                              & $P_{f_{NL}-BOX}$ & $\frac{7}{8}$              \\ \hline
\multirow{5}{*}{$f_{CTC}$}    & $P_{WN}$         & $\frac{1}{2}$              \\ \cline{2-3}
                              & $P_{LV}$         & $\frac{1}{2}$              \\ \cline{2-3}
                              & $P_{SINGLET}$    & $\frac{1}{2}$              \\ \cline{2-3}
                              & $P_{PR-BOX}$     & $\frac{1}{2}$              \\ \cline{2-3}
                              & $P_{f_{NL}-BOX}$ & $\frac{1}{2}$              \\ \hline
\end{tabular}
\caption{In this Table we enlist down the respective efficiency factor for post selection functions 
$f$: $f_{sig1}$, $f_{sig2}$, $f_{sig}$, $f_{B-CHSH}$, $f_{CTC}$
and a given input probability distributions $P_{in}$: $P_{WN}$, $P_{LV}$, $P_{SINGLET}$, $P_{PR-BOX}$, $P_{f_{NL}-BOX}$ }
\end{table}

\noindent One can notice that such post-selection are fairly costly. 
Post-selection in real world does not alter the underlying probability distribution. 
As a consequence there is no violation of the Pigeon hole principle in the classical world. 
While it is not known yet whether $RP = NP$, it is however interesting to note why the technique employed here 
does not suffice to prove it. But, even with the given construction, one would require an expected $\frac{1}{Pr(P=1)}$, which is exponential, runs of the machine to get a selective run. Guessing boolean assignments at random and then verifying whether the formula 
satisfies it, has a probability of success, in single run, $\frac{s}{2^n}$. Thus, for such a method to have a 
probability greater than $\frac{1}{2}$, one would have to repeat the experiment exponential number of times.

\noindent Next we re discuss two important properties of post selection function namely orthogonality and Boolean 
compliments in terms of efficiency factor $\eta^{P_{in}}_f$.\\
 
\noindent \textbf{a) Sequential application and (semi-)orthogonal functions.}\\

\noindent Start again with $P_{WN}$, the drop in efficiency on sequential application of two 
functions $f$ (say first) and (then) $f_1$ are given by,
\begin{equation}
\eta^{P_{WN}}_{f,f^1}=P_{WN}((f=0)\wedge (f^1=0)).
\end{equation}
\noindent If $f$ and $f_1$ are orthogonal then,
\begin{equation}
\eta^{P_{WN}}_{f,f^1}=0.
\end{equation}

\noindent \textit{Definition 12.}
\noindent Two functions are semi orthogonal if,
\begin{equation}
\eta^{P_{WN}}_{f,f^1}<\eta^{P_{WN}}_{f}\eta^{P_{f_{BOX}}}_{f^1}.
\end{equation}

\noindent \textit{Definition 13.}
\noindent Two functions are $f$ and $f^1$ are non orthogonal if
\begin{equation}
\eta^{P_{WN}}_{f,f^1}=\eta^{P_{WN}}_{f}\eta^{P_{f_{BOX}}}_{f^1},
\end{equation}
\noindent which is the case with $f_2$ and $f_{final}$.\\

\noindent \textbf{b) Boolean Compliments.}\\

\noindent If two functions $f$ and $f^1$are Boolean compliments that is $f=f^1\oplus 1$  then,
\begin{enumerate}
\item: As $P((f=0)\wedge (f^1=0))=0$,
\begin{equation}
\eta^{P_{WN}}_{f,f^1}=0.
\end{equation}
\item As $P((f=0)\vee (f^1=0))=1$
\begin{equation}
\eta^{P_{WN}}_{f}+\eta^{P_{WN}}_{f^1}=1.
\end{equation}

\end{enumerate}

\noindent We can simultaneously apply $f$ and $f^1$  as in principle we could have two PSD applied to $P_{WN}$ such that 
one accepts when $f=0$ and the other when $f=1$.

\subsection{ From an adversarial perspective. }

\noindent From an adversarial perspective, faking correlations, in particular Bell violation is of great importance. 
The fact that Eve cannot fake (simulate) non-local correlations (at $\eta=1$ using post-selection leads 
to device independently secure self assessment, QKD (Quantum Key Distribution scheme), randomness expansion and  so on. 
However at lower efficiency a  Eve could apply post-selection (denial of service attack) and fake 
correlations ( Bell violation in particular ). We provide  a (optimal) protocol for potential Eves dropper and 
study the relationship between input/output (actual/apparent)probability distribution and the device independent 
efficiency factor associated with them. As a result with provide robust bounds on minimum efficiency for non-locality 
of singlet statistics and for $\epsilon$ bell violation.\\

\noindent In general lets say Eve starts with $P_{in}$ with some 
$P_{in}(f=0)$ and wants to simulate $P_{out}(f=0)>P_{in}(f=0)$. She can do this by following the protocol,

\begin{enumerate}
\item Whenever $f=0$ accept the trial.
\item Whenever $f=1$, with $p\in \{0,1\}$ probability accept the trial.
\end{enumerate}
\noindent Here the efficiency $\eta^{P_{WN}}_{x.y\oplus u \oplus v}=\frac{1+p}{2}$, so $P_{out}(f=0)=\frac{1}{1+p}$. 
So Eve can cheat Alice and Bob to believe that they share a correlation with $P_{out}(f=0)$ at maximum efficiency,  
\begin{equation}
\eta^{P_{in}}_{f}=\frac{P_{in}(f=0)}{P_{out}(f=0)}.
\end{equation}

\noindent The malicious Eve wants to simulate the statistics of the singlet quantum state in a Bell experiment. 
We already know it is impossible to do this at $\eta^{P^{in}}_f=1$ or the case of perfect (detectors) devices. 
However at lower $\eta^{P^{in}}_f\leq1$  it possible to apply quantum Bell violation. 
So Eve can cheat Alice and Bob to believe that they share a singlet state with an efficiency factor at most equal to
 \begin{equation}
\eta^{P^{WN}}_{x.y\oplus u \oplus v}=\frac{1}{2max_Q(P^{out}_{B-CHSH})}=\frac{2}{2+\sqrt[]{2}}=0.585786.
\end{equation}
\noindent $P_{WN}$ cannot simulate singlet statistics at efficiency above $\eta^{P^{WN}}_{x.y\oplus u \oplus v}$. 
How ever Eve could use other classical correlations such as $P_{in}=max(P^{B-CHSH}_{LV})=\frac{3}{4}$. 
She follows the same protocol. Now  the efficiency $\eta^{P_{in}}_{x.y\oplus u \oplus v}=\frac{3+p}{4}$, 
so $P^{out}_{B-CHSH}=\frac{3}{3+p}$.  So Eve can cheat Alice and Bob to believe that they share a singlet state with 
\begin{equation}
\eta^{P_{LV}}_{x.y\oplus u \oplus v}=\frac{3}{4max_Q(P^{out}_{B-CHSH})}=\frac{3}{2+\sqrt[]{2}}=0.87867 
\end{equation}
$P_{LV}$ cannot simulate singlet statistics at efficiency above $\eta^{P^{LV}}_{x.y\oplus u \oplus v}$, so the 
singlet statistics can guarantee Bell-Violation at higher efficiency.
In general for $\epsilon \in(0,\frac{1}{4})$ one requires,
\begin{equation}
\eta^{P_{LV}}_{x.y\oplus u \oplus v}=\frac{\frac{3}{4}}{\frac{3}{4}+\epsilon}.
\end{equation}

\noindent In TABLE XVI we write down the bounds of the efficiency factor  $\eta^{P_{in}}_f$ for a given input probability 
distribution $P_{in}$, post selection function $f$ and the output probability distribution $P_{out}$.\\

\begin{table}[h]
\begin{tabular}{|c|c|l|c|c|}
\hline
\textbf{$f$}                  & \textbf{$P_{in}$}           & \textbf{$P_{out}$} & \textbf{$P_{out}(f_{B-CHSH}=0)$} & \textbf{$\eta^{P_{in}}_f$} \\ \hline
\multirow{4}{*}{$f_{B-CHSH}$} & \multirow{2}{*}{$P_{WN}$} & $P_{SINGLET}$      & $0.85355339059$                & $0.585786$                 \\ \cline{3-5} 
                              &                           & $P_{PR-BOX}$       & $1$                            & $\frac{1}{2}$              \\ \cline{2-5} 
                              & \multirow{2}{*}{$P_{LV}$} &   $P_{SINGLET}$     & $0.85355339059$                & $0.87867$                  \\ \cline{3-5} 
                              &                           &  $P_{PR-BOX}$     & $1$                            & $\frac{3}{4}$              \\ \hline
\end{tabular}
\caption{Bounds on $\eta^{P_{in}}_f$ for input probability distribution $P_{in}$, output probability distribution $P_{out}$  
and post selection function $f$.}
\end{table}

%\section{Conclusion} In this work we have considered

%\noindent \subsection{no-security at $\frac{1}{2}$ efficiency}
%We saw above that there is no secure Bell violation at $\eta=\frac{1}{2}$ rendering the entanglement based useless. 
%Also in prepare and measure scenario, MAKAROV attack on imperfect detectors with $\eta=\frac{1}{2}$ renders the 
%protocol insecure, as $P(b=e)=\frac{1}{2}$. However in a separate work we show that Semi Device Independent 
%security is possible at any $\eta>\frac{1}{2}$.  
\bibliographystyle{ieeetr}

\bibliography{cite}

\end{document}